\pdfoutput=1
\documentclass[two column,prl,showpacs,amsmath,amssymb,superscriptaddress]{revtex4}
\usepackage{bm}
\usepackage{graphicx}
\usepackage{amssymb}
\usepackage{xcolor}

\begin{document}

\title{Schottky anomaly and short-range antiferromagnetic correlations in filled skutterudites Pr$_{1-x}$Eu$_x$Pt$_4$Ge$_{12}$  }

\author{R. B. Adhikari}  
\affiliation{Department of Physics, Kent State University, Kent, Ohio, 44242, USA}

\author{P. Shen}  
\affiliation{Department of Physics, Kent State University, Kent, Ohio, 44242, USA}

\author{D. L. Kunwar}  
\affiliation{Department of Physics, Kent State University, Kent, Ohio, 44242, USA}

\author{I. Jeon}
\affiliation{Center for Advanced Nanoscience, University of California, San Diego, La Jolla, California 92093, USA}
\affiliation{Materials Science and Engineering Program, University of California, San Diego, La Jolla, California 92093, USA}

\author{M. B. Maple}
\affiliation{Center for Advanced Nanoscience, University of California, San Diego, La Jolla, California 92093, USA}
\affiliation{Materials Science and Engineering Program, University of California, San Diego, La Jolla, California 92093, USA}
\affiliation{Department of Physics, University of California at San Diego, La Jolla, CA 92903, USA}

\author{M. Dzero}
\affiliation{Department of Physics, Kent State University, Kent, Ohio, 44242, USA}

\author{C. C. Almasan}
\affiliation{Department of Physics, Kent State University, Kent, Ohio, 44242, USA}

\date{\today}
\pacs{71.10.Hf, 71.27.+a, 74.70.Tx}

\begin{abstract}
By performing a series of thermodynamic measurements in an applied magnetic field $H_{\textrm{ext}}$, we investigated the effects of Eu substitution on the Pr sites in filled skutterudite compound Pr$_{1-x}$Eu$_x$Pt$_4$Ge$_{12}$ ($ 0 \leq x \leq 1$). A heat capacity Schottky anomaly is present over the whole doping range. For the samples with $x > 0.5$, these Schottky anomaly peaks shift to lower temperature with increasing $H_{\textrm{ext}}$. We argue that this behavior reflects the antiferromagnetic (AFM) ordering of the Eu moments, as the AFM transition is suppressed by $H_{\textrm{ext}}$. The Schottky peaks in the samples with $x \leq 0.5$ shift to higher temperatures with increasing magnetic field, signaling the presence of an internal magnetic field due to short-range AFM correlations induced by magnetic moments of neighboring Eu sites. In low $H_{\textrm{ext}}$, the Schottky gaps show a non-linear relationship with $H_{\textrm{ext}}$ as the magnetic moments become weakly magnetized. In high $H_{\textrm{ext}}$, the magnetic moments of Eu sites become completely aligned with $H_{\textrm{ext}}$. Thus, increasing $H_{\textrm{ext}}$ does not further increase the magnetization, hence the Schottky gaps increase linearly with $H_{\textrm{ext}}$.
\end{abstract}

\pacs{71.10.Ay, 74.25.F-, 74.62.Bf, 75.20.Hr}

\maketitle
\section{Introduction}

Filled skutterudite compounds with the chemical formula $M$Pt$_4$Ge$_{12}$ (where $M$ denotes alkaline earth, lanthanide, or actinide) belong to the family of heavy-fermion superconductors. These materials have attained a renewed experimental and theoretical interest recently due to their potential to host unconventional electronic many-body ground states, such as various multipolar orders which are driven by magnetic degrees of freedom at low temperatures. A competition and/or possible coexistence between unconventional superconductivity and various - possibly nontrivial - magnetic ground states provides a major motivation for further exploring the physics of these compounds.  

PrPt$_4$Ge$_{12}$ is a heavy-fermion superconductor which has a moderately high - relative to other heavy-fermion superconductors - critical temperature of $T_c\simeq{7.9}$ K and smaller $\gamma\sim 60$ mJ/(mol$\cdot$K$^2$), corresponding to a medium enhancement of the conduction electron effective mass~\cite{Maisuradze2009}. It is important to keep in mind that PrPt$_4$Ge$_{12}$ is also multiband superconductor with two Fermi surfaces having one nodal and one nodeless gap indicating the unconventional nature of superconductivity (SC). Specifically, its multiband nature may be the main reason why superconductivity remains fairly robust with respect to introducing disorder by chemical substitutions. Indeed, the substitution of Ce to the filler cites of Pr leads to the overall suppression of $T_c$ with the nodal gap being gradually suppressed as well \cite{Singh2016}. Generally, the unconventional SC in Pr-based scutterudites  is likely driven by system's proximity to magnetic ground states. In fact, several Eu-based compounds have been found to exhibit antiferromagnetism (AFM) \cite{Ren2008,Midya2016}. 

Compounds containing Eu$^{2+}$ (or Gd$^{3+}$) ions have large total angular momentum $J=S=7/2$ per Eu (Gd) and hence are expected to host large magnetic moments. Consequently, these compounds typically exhibit magnetic ordering. Some of the filled skutterudite systems with Eu$^{2+}$ electronic configuration such as  EuFe$_4$Sb$_{12}$ and  EuFe$_4$As$_{12}$, indeed, show magnetic ordering at  temperatures  $T_C\sim$ 88 K and $\sim$ 152 K, respectively, where the higher $T_C$ has been
attributed to the existence of an additional, {albeit} small, magnetic moment ($\sim$ 0.21$\mu_B$ for the Fe-Sb cage) on the Fe ion \cite{Bauer2004,Sekine2009}. Furthermore,  EuPt$_4$Ge$_{12}$ has AFM with the N\'{e}el critical temperature $T_N\sim$ 1.78 K, an effective magnetic moment of $\mu_{\textrm{eff}}=7.4\mu_B$, and a Curie-Weiss temperature $\Theta_{CW}\sim 
-11$ \nolinebreak K \cite{Nicklas2011,Grytsiv2008}. 
The lower magnetic ordering temperature for EuPt$_4$Ge$_{12}$ may be due to strong fluctuations and the absence of
a magnetic moment on Pt in the Pt-Ge cage, causing a decrease in the exchange coupling determined by the Ruderman-Kittel-Kasuya-Yosida (RKKY) interaction between the Eu$^{2+}$ localized magnetic moments and
the conduction-electron spins \cite{Grytsiv2008,Krishnamurthy2007}.

Heat capacity measurements of Eu- or Gd-based samples have been found to exhibit upturn in $C_e/T$ upon lowering the temperature. This upturn happens to be due to Schottky anomaly resulting from the splitting of the ground state octet of Eu/Gd by the internal molecular and externally applied magnetic field \cite{Naugle2006,Nigel2003,Kohler2006}. Further studies have shown upturns in $C_e/T$ vs $T$ data containing Pr and have been attributed to crystalline electric field (CEF) splitting of the ground state of Pr ions \cite{Frederick2005, Takeda2000}. Thus, at least in principle, it should be possible to probe the signatures of magnetic correlations by
analyzing the Schottky contribution to the heat capacity, since the CEF structure of the Pr ion is well known.

 We performed low-temperature specific heat  measurements on  samples of Pr$_{1-x}$Eu$_x$Pt$_4$Ge$_{12}$ in magnetic field. Our detailed and systematic analysis indicates that the upturns in the heat capacity are caused by the splitting of the octet degenerate states of Eu$^{2+}$ due to the internal magnetic field that co-exists with superconductivity \cite{Adhikari2018}. We have also systematically analyzed  the effect of magnetic field on the temperature where the Schottky peaks appear. Our analysis shows that the temperature where the Schottky peaks appears increases linearly with increasing magnetic field in the high Eu content and high magnetic field region. For the low Eu-substituted samples and in the low magnetic field region, the temperature of the Schottky peak show a non-linear magnetic field dependence since more and more Eu magnetic moments become weakly magnetized with increasing applied field. 

\section{Experimental Details}
Pr$_{1-x}$Eu$_x$Pt$_4$Ge$_{12}$ samples  were synthesized by arc-melting and annealing from high purity Pr ignots, Eu ignots, Pt sponge, and Ge pieces according to the procedure described in detail in Ref. \cite{Ijeon2016}. The crystal structure was determined through X-ray powder diffraction using a Bruker D8 Discover X-ray diffractometer with Cu-K$_{\alpha}$ radiation, and the XRD patterns were analysed through Rietveld refinement \cite{Toby2000}. 
A detailed sample characterization of the series of these polycrystalline samples through X-ray diffraction, electrical resistivity, and magnetic susceptibility, as described elsewhere \cite{Ijeon2017}, shows the purity of the samples used in this study. 

The two surfaces of each sample were polished with sand paper  to improve the contact between the sample and the specific heat platform. We performed a series of specific heat measurements on the polycrystalline samples of Pr$_{1-x}$Eu$_x$Pt$_4$Ge$_{12}$ sith $x =$ 0, 0.05, 0.10, 0.15, 0.20,  0.30,  0.38,  0.50, 0.70, 0.80, 0.90, and 1.00 in applied magnetic field $H_{\textrm{ext}}$ up to 14 T over the temperature $T$ range $0.50$ K $\leq T \leq 10$ K. The specific heat  measurements  were  performed via a standard thermal relaxation technique using the He-3 option of a Quantum Design Physical Property Measurement System (PPMS).

\section{Results and Discussion}

\begin{figure}
\centering
\includegraphics[width=1.0\linewidth]{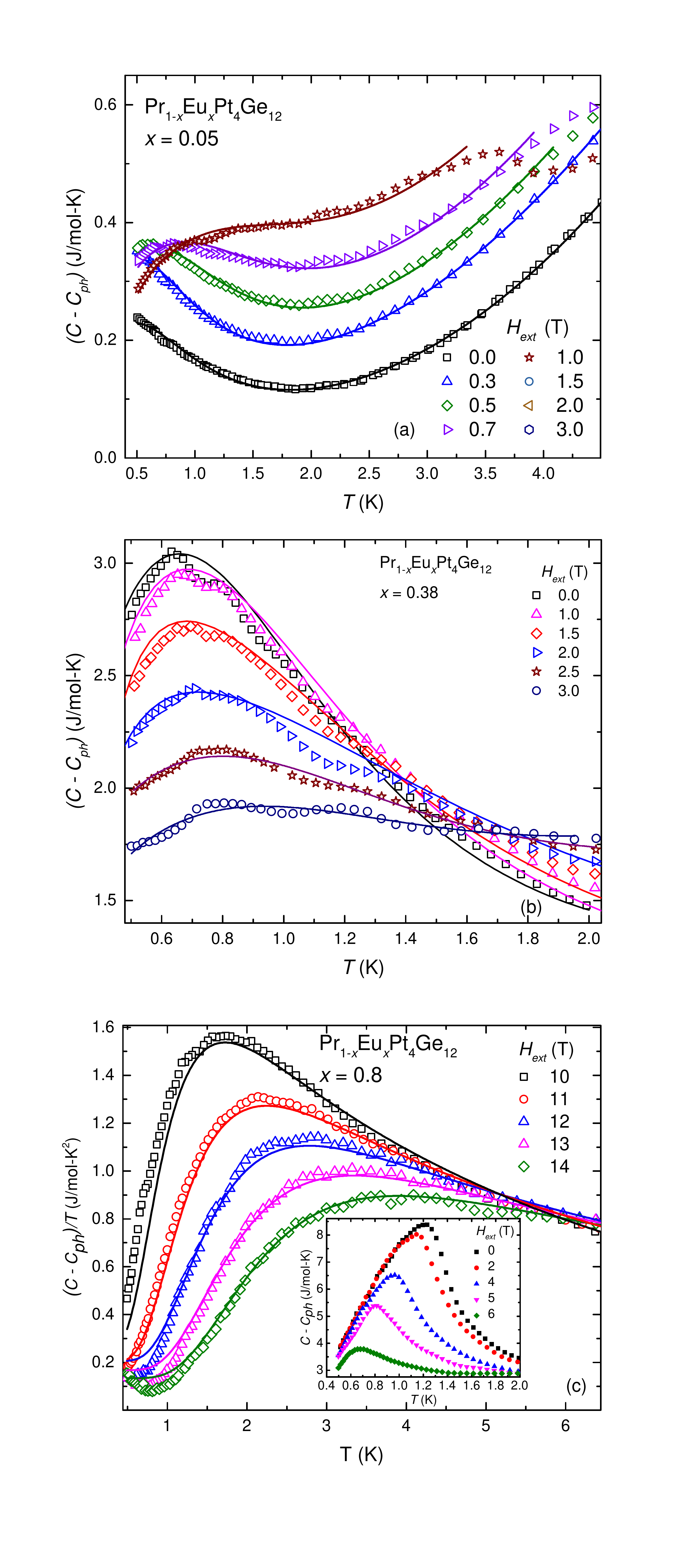}
\caption{(Color online) Specific heat $C-C_{ph}$ vs temperature $T$ of Pr$_{1-x}$Eu$_x$Pt$_4$Ge$_{12}$ measured in different magnetic fields $H$ for the (a) $x=0.05$, (b) $x=0.38$, and (c) $x=0.80$ samples. The solid curves are fits of the data using the sum of  Schottky and superconducting contributions for (a) and (b), and Schottky and normal-state contributions for (c), as described in the text.}
\label{Fig1}
\end{figure}

The measured specific heat in the normal state $C(T)$ = $\gamma_n T$+$B T^3$ is the sum of electronic 
$C_{\rm e}\equiv \gamma_n T$ ($\gamma_n$ is the normal-state Sommerfeld coefficient) and phonon $C_{\textrm{ph}} \equiv B T^3$ contributions; hence, we did a least-square fit of $C/T$ vs $T^2$ data in the normal state ($T_c < T \leq15$ K) for different Eu concentrations, as described and shown in Fig. 1 of ref. \cite{Adhikari2018},  in order to determine $\gamma_n$ and $B$. We then subtracted the phonon contribution to the specific heat for all the measured samples of Pr$_{1-x}$Eu$_x$Pt$_4$Ge$_{12}$. We subtracted the same phonon contribution from the specific heat data measured in an applied magnetic $H_{\textrm{ext}}$, thus assuming that the phonon contribution is field independent. 

Figures~\ref{Fig1}(a), \ref{Fig1}(b), and  \ref{Fig1}(c) display the $C-C_{ph}$ vs $T$ data for the $x=0.05$, 0.38, and 0.80, respectively, samples measured in different applied magnetic fields $H_{\textrm{ext}}$. These figures reveal upturns present in the specific heat data below 2 K. For the samples with $x\leq0.15$, the specific heat shows an upturn without reaching a maximum [Fig.~\ref{Fig1}(a)], whereas the samples with larger $x$ values show a clear peak present in the low-temperature region [Figs.~\ref{Fig1}(b) and \ref{Fig1}(c)]. 

With increasing applied magnetic field $H_{\textrm{ext}}$, the samples with $0.05\leq x \leq 0.15$ begin to show a distinct peak [Fig.~\ref{Fig1}(a)], while the samples with  $x \leq 0.5$ show that the peak becomes broader, shifts to higher temperatures, and decreases in amplitude [Fig.~\ref{Fig1}(b)]. 
The Schottky peaks for the $0.7\leq x \leq 1$ samples reveal a long-range AFM transition at $T_N$, see inset Fig. \ref{Fig1}(c), that, as expected,  shift to lower temperatures with increasing $H_{\textrm{ext}}$. Nevertheless, once the AFM transition is suppressed below the lowest measured temperature of  0.5 K, the Schottky peak shifts to higher temperatures with further increase $H$ [Fig. \ref{Fig1}(c)], as also observed in the $x < 0.5$ samples [Figs.~\ref{Fig1}(a) and \ref{Fig1}(b)].      

We note that a superconducting transition at a temperature $T_c$ is clearly seen in the heat capacity of the  samples with $0\leq x \leq 0.3$ \cite{Adhikari2018}. The superconducting jump becomes broader and $T_c$ shifts to lower temperatures, as expected, with increasing $H_{\textrm{ext}}$ and increasing Eu concentration \cite{Adhikari2018}. The samples with $0.3 < x \leq 0.5$ show a superconducting transition only in resistivity measurements \cite{Ijeon2017}, while the samples with $x > 0.5$ do not display a superconducting transition for temperatures down to $T = 0.5$ K.

\begin{figure}
\centering
\includegraphics[width=\linewidth]{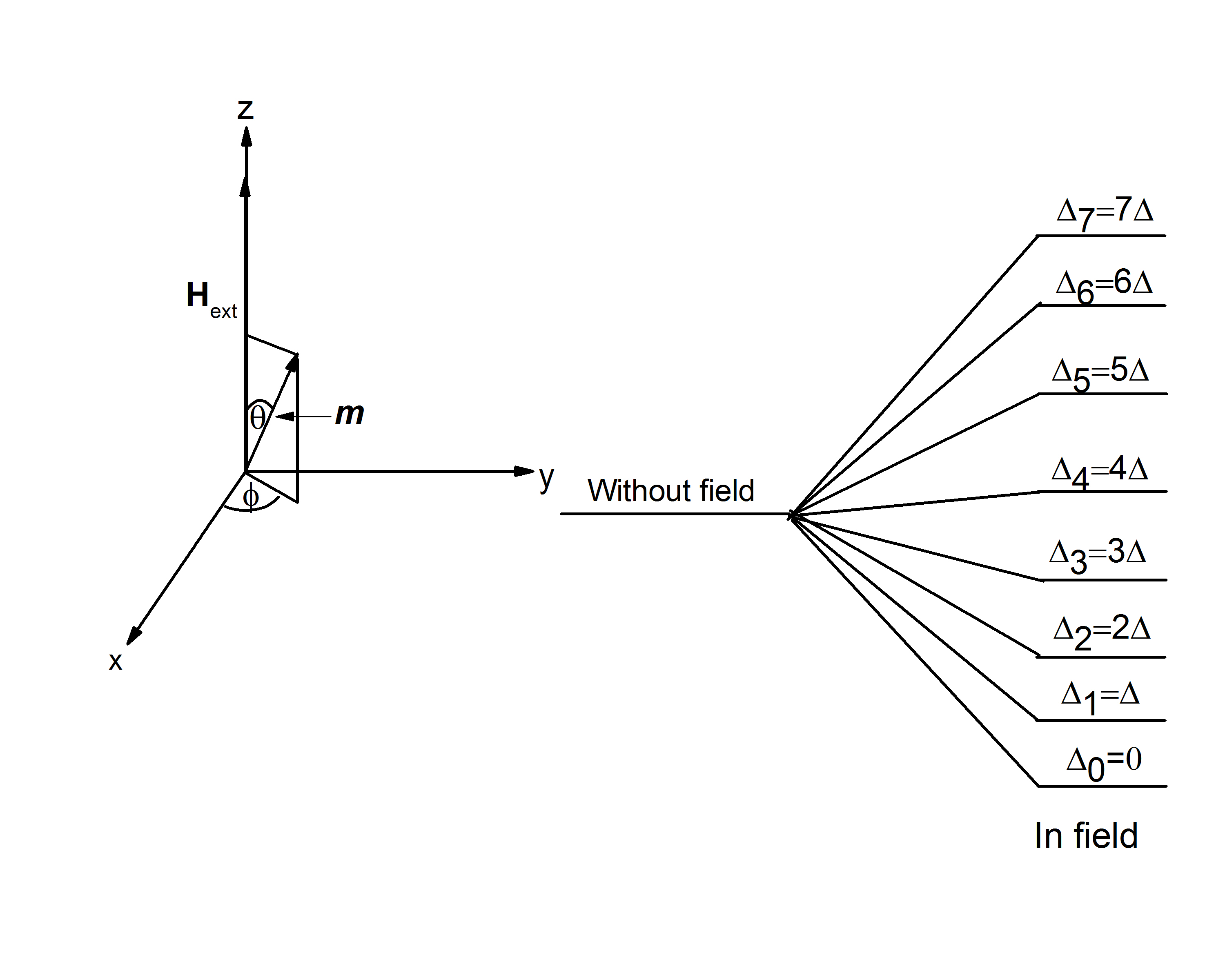}
\caption{(Color online) Left panel: Schematic representation of the net magnetic moment $\textbf{m}$  and external magnetic field pointing in different directions. Right panel: Splitting of the ground state $^8S_{7/2}$ of Eu into 8 equally-spaced energy levels by both internal and applied magnetic fields. The lines are not to a scale.}
\label{Fig2}
\end{figure}

We have attributed the Schottky anomaly present in the heat capacity data of the $x \leq 0.5$ samples in $H_{\textrm{ext}}$ to the splitting of the degenerate ground state of Eu$^{2+}$ into eight equally-spaced energy levels, as shown schematically in Fig.~\ref{Fig2}, by the internal field $H_{\textrm{int}}$ due to the net magnetic moment $\textbf{m}$ present as a result of short-range antiferromagnetic correlations between the nearest-neighbor Eu ions \cite{Adhikari2018}. These short-range antiferromagnetic correlations co-exist with superconductivity in these lower Eu substituted samples (see Fig. 7 of Ref. \cite{Adhikari2018}). 
We note that one may expect short-range antiferromagnetic correlations between the Eu ions to be present in the alloys Pr$_{1-x}$Eu$_x$Pt$_4$Ge$_{12}$ since EuPt$_4$Ge$_{12}$ orders antiferromagnetically.  

The Schottky heat capacity anomaly for an eight-energy-levels system with the degeneracy fully lifted by magnetic field is given by \cite{Gopal1966} 
\begin{equation}\label{Sch}
\begin{split}
C_{\textrm{Sch}}&=r(x)\frac{R}{T^2}\left[\frac {f_2(T)}{f_0(T)}-\frac {f_1^2(T)}{f_0^2(T)}\right], \\
f_m(T)&=\sum\limits_{j=0}^7\Delta_j^m\exp\left(-\frac{\Delta_j}{k_BT}\right), m=0,1,2 
\end{split}
\end{equation}
where $r(x)$ is one of the fitting parameters and it turns out that it represents the concentration of Eu ions \cite{Adhikari2018}, $R=8.31$ J/mol-K is the universal gas constant, $\Delta_j=j\cdot\Delta$ is the energy  gap between the lowest energy level ($j=0$) and the $j^{th}$ energy level, $\Delta \equiv g\mu_BH_{\textrm{eff}}/k_B$, $g=2$ ($L=0$), $\mu_B$ is the Bohr magnetron, and $H_{\textrm{eff}}$ is the effective magnetic  field. Hence, the value of the Schottky gap $\Delta$ fully depends on $H_{\textrm{eff}}$. In the presence of an applied magnetic field $H_{\textrm{ext}}$, both $H_{\textrm{int}}$ and $H_{\textrm{ext}}$ split the degenerate $^8S_{7/2}$ ground state of Eu into 8 equally-spaced energy levels.
Hence, the effective field has two contributions: $\textbf H_{\text{eff}}\equiv\textbf H_{\textrm{int}}+\textbf H_{\textrm{ext}}$.  

As just mentioned, the lower Eu-doped samples ($0.05\leq x\leq 0.5$) have two contributions to the heat capacity in the low field and low temperature ($0.5 \leq T<T_c$) region: Schottky and superconducting contributions. Specifically, for the samples with Eu  concentrations in the range $0.05\leq x \leq 0.15$, the superconducting contribution is best described by $(C-C_{ph})\propto T^2$, i.e., line nodes in the superconducting gap, while for $0.2\leq x \leq 0.5$ by $(C-C_{ph})\propto \exp^{-\delta/T}$, i.e., by an isotropic gap; this is perfectly expected since the nodal gap is quickly suppressed by scattering on lattice imperfections. In the high field region where $T_c<0.5$ K or for the $x \geq 0.7$ samples, the total specific heat is the sum of Schottky and normal-state electronic ($\gamma_nT$) contributions. Least-square fits of the data are shown by the solid curves in Figs.~\ref{Fig1}(a) through \ref{Fig1}(c). The fitted curves are in excellent agreement with the measured specific heat data. 

\begin{figure}
\centering
\includegraphics[width=\linewidth]{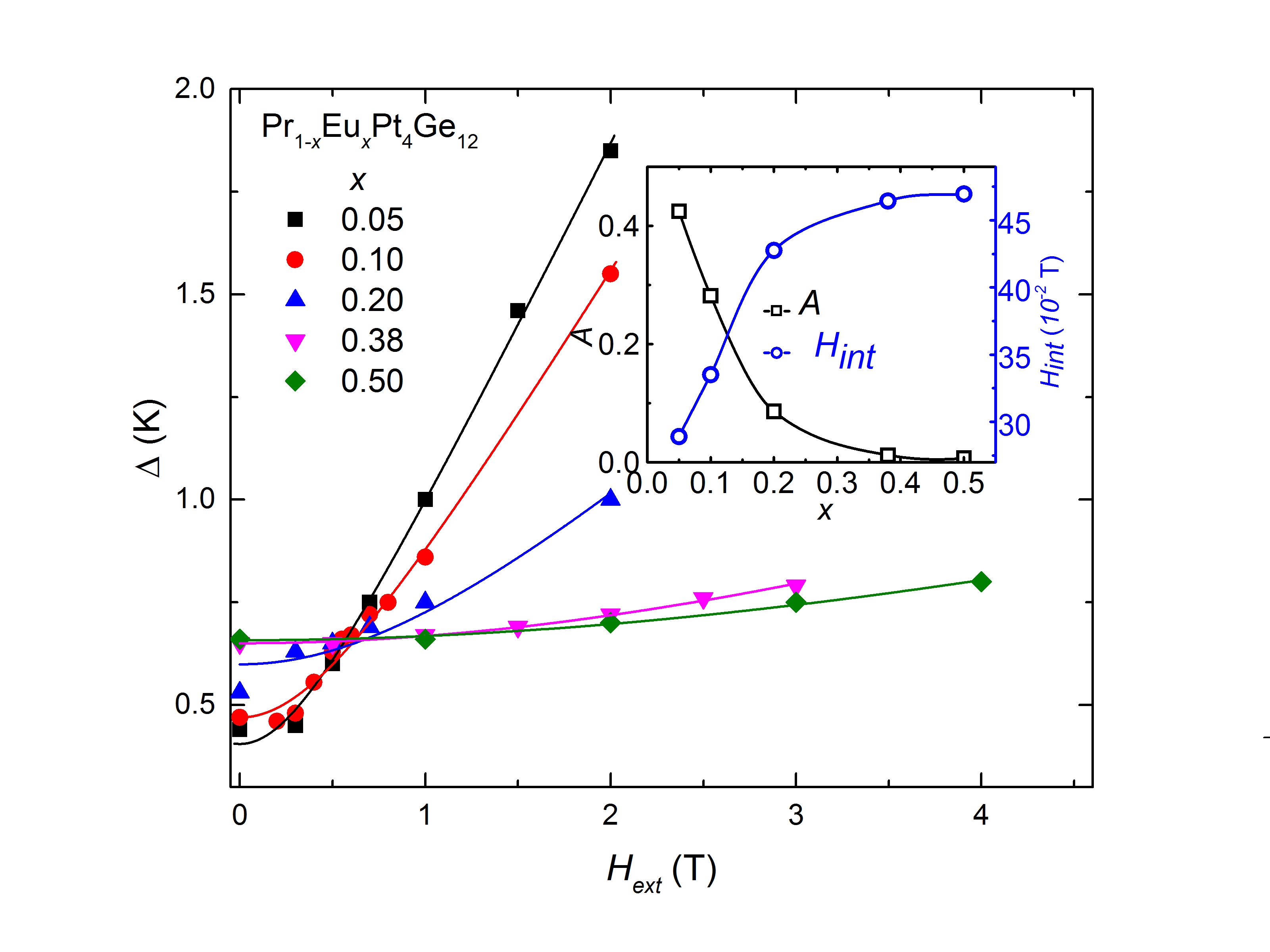}
\caption{(Color online) Plots of  the Schottky gap $\Delta$ vs $H_{\textrm{ext}}$, for relatively small applied magnetic fields. The  solid curves are fits of the data as discussed in the text. Inset: Fitting parameters $A$ (left vertical axis) and $H_{int}$ (right vertical axis) plotted as a function of Eu-substitution $x$.}
\label{Fig3}
\end{figure}

\begin{figure}
\centering
\includegraphics[width=1\linewidth]{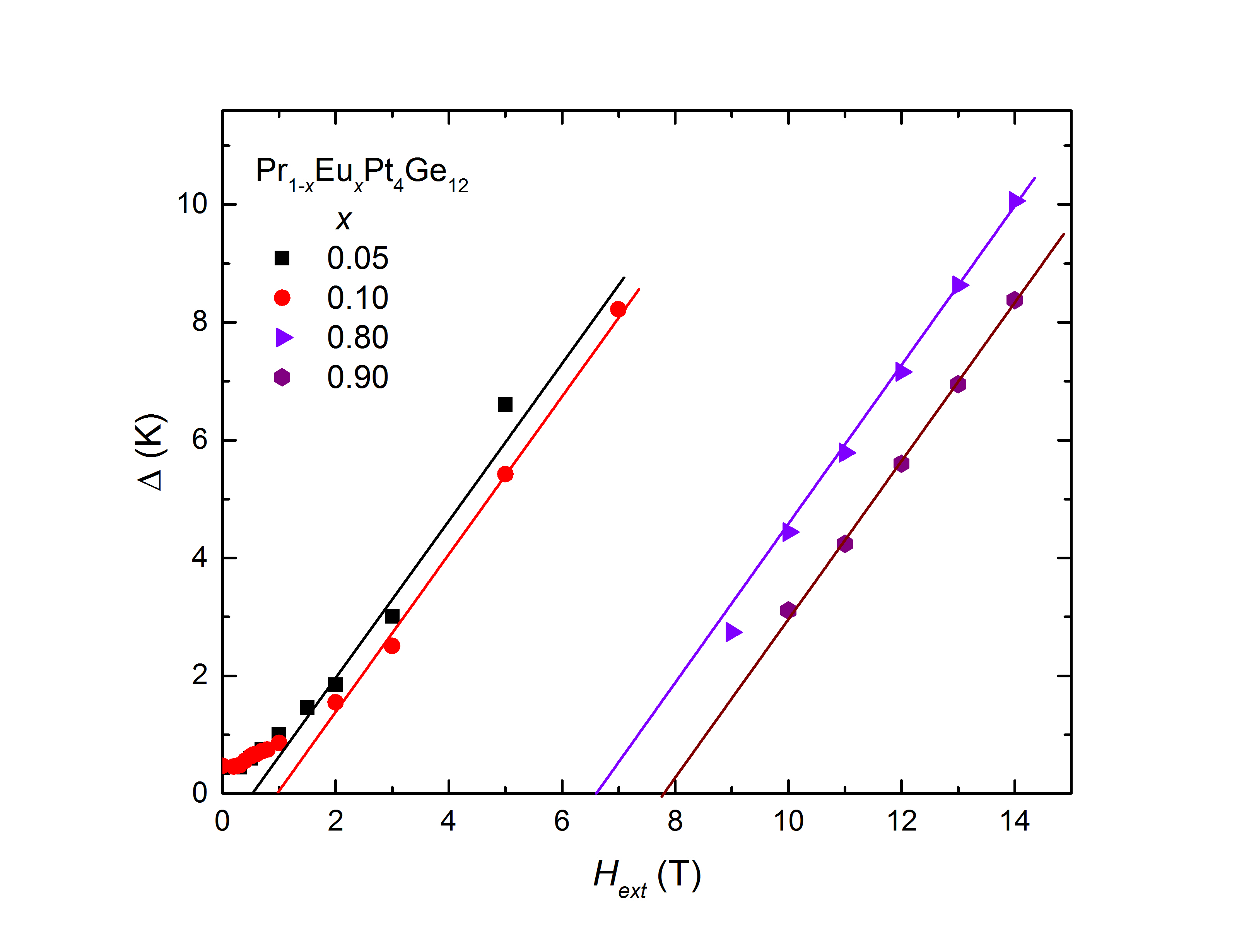}
\caption{(Color online) Plots of $\Delta$ vs $H$ data for high applied magnetic fields. The straight lines are guides to the eye and show that all the $\Delta$ vs $H$ plots have same slope of ($1.4 \pm0.04$) K/T.}
\label{Fig4}
\end{figure}

The values of the Schottky gaps obtained from these fits are plotted as a function of applied magnetic field $H_{\textrm{ext}}$ for various Eu substitutions  in Figs.~\ref{Fig3} and ~\ref{Fig4}. We note that the Schottky gap $\Delta$ for a system with eight energy levels and the degeneracy fully lifted is also given by the temperature corresponding to the Schottky peak, i.e., $\Delta=T_{\textrm{peak}}$. The values of $ \Delta$ obtained from fits (as discussed above) and from $T_{\textrm{peak}}$ for the samples for which the Schottky peak is clearly observed are in excellent agreement. Hence, some $\Delta$ values shown in these figures for some $x$ and/or  $H_{\textrm{ext}}$ values were also extracted from $T_{\textrm{peak}}$. Notice that the Schottky gaps are non-linear in  $H_{\textrm{ext}}$ for small values of  $H_{\textrm{ext}}$ and increase linearly with $H_{\textrm{ext}}$ with a doping-independent slope for large values of $H_{\textrm{ext}}$.
  
We gained an understanding of the physics that governs the behavior of these materials in the presence of an applied magnetic field as follows. Taking $\textbf H_{\textrm{ext}}=H_{\textrm{ext}}\hat z$, and witting $\textbf H_{\textrm{int}}= H_{x} \hat x+H_{y} \hat x+H_{z} \hat z$, gives $\langle\textbf H^2_{\text{eff}}\rangle=\langle H^2_x+H_y^2+(H_{\textrm{ext}}+H_z)^2\rangle$. To obtain a function that describes the effective field at intermediate field values, we treat the magnetic degrees of freedom as classical, freely-rotating magnetic moments $\textbf{m}$.
The energy of the magnetic moment is $E=-m H_{\textrm{ext}} \cos\theta$, where $\theta$ is the angle between $\textbf H_{\textrm{ext}}$ and $\textbf m$ with $\textbf H_{int} = - \textbf{m}$. The partition function is then:
 
\begin{equation} 
 Z=\int_0^{2\pi}d\phi\int_0^\pi d\theta \sin\theta e^{ \beta m H_{\textrm{ext}} \cos\theta}=4\pi \frac{\sinh \beta m H_{\textrm{ext}}}{\beta m H_{\textrm{ext}}}.
\label{Partition function}
\end{equation}
 The expectation value of $\textbf H^2_{\text{eff}}$ is
 \begin{equation} 
 \begin{aligned}
\langle\textbf H^2_{\text{eff}}\rangle=&\frac 1 Z \int_0^{2\pi}d\phi\int_0^\pi d\theta \sin e^{ \beta m H_{\textrm{ext}} \cos\theta}\times  \\&[(H_{\textrm{ext}}-H_{\textrm{int}}(x)\cos\theta)^2+(H_{\textrm{int}}(x)\sin\theta)^2]\\
=&H_{\textrm{ext}}^2+H_{\textrm{int}}^2(x)-2H_{\textrm{ext}}H_{\textrm{int}}(x) L(\beta mH_{\textrm{ext}}),
\end{aligned}
\label{Heff}
\end{equation}
 where the Langevin function is $L(\beta mH_{\textrm{ext}})=\coth(\beta mH_{\textrm{ext}})-1/(\beta mH_{\textrm{ext}})$ and $H_{\textrm{int}}(x)$ is proportional to the concentration of Eu atoms. In the limit $H_{\textrm{ext}}\to0$, we find $\langle H_{\text{eff}}^2\rangle\approx H_{\textrm{int}}^2(x)$, so for $\beta mH_{\textrm{ext}}\ll 1$ it follows that $L(\beta mH_{\textrm{ext}})\approx (\beta mH_{\textrm{ext}})
 /3$ and 
$ \langle H_{\text{eff}}^2\rangle=H_{\textrm{ext}}^2-2/3H^2_{\textrm{ext}}H_{\textrm{int}}(x)\beta m+H_{\textrm{int}}^2(x)$. 
 $\Delta \equiv g\mu_BH_{\textrm{eff}}/k_B$ becomes
\begin{equation}
\Delta=(g\mu_B/k_B)\sqrt{A(x)H_{\textrm{ext}}^2+H_{\textrm{int}}^2(x)},
\label{Delta2}
\end{equation}
 where $A(x)=1-2\beta mH_{\textrm{int}}(x)/3$. 
In the limit $\beta mH_{\textrm{ext}}\gg 1$, $L(\beta mH_{\textrm{ext}})=1$ and 
$\langle H_{\text{eff}}^2\rangle=\left(H_{\textrm{ext}}-H_{\textrm{int}}(x)\right)^2$. Hence,
\begin{equation}
\Delta=(g\mu_B/k_B)\left(H_{\textrm{ext}}-H_{\textrm{int}}(x)\right).
\label{Delta}
\end{equation}
  
We fitted the $\Delta(x)$ vs $H_{\textrm{ext}}$ data of Figs. ~\ref{Fig3} and ~\ref{Fig4} with Eqs. \ref{Delta2} and \ref{Delta}, respectively, with $A(x)$ and $H_{\textrm{int}}(x)$ as fitting parameters. Notice the excellent fits obtained especially for the non-linear $\Delta(x)$ vs $H_{\textrm{ext}}$ data of Fig. ~\ref{Fig3}. When the externally applied magnetic field is large enough, it aligns the net magnetic moment  in its direction, hence, $ \Delta$ becomes proportional to the resultant magnetic field. In this high field region where the superconducting contribution of the $0.05\leq x\leq0.5$ samples or the antiferromagnetic contribution of the $0.7\leq x\leq0.9$ samples is suppressed below 0.5 K, $ \Delta$ increases linearly with $H_{\textrm{ext}}$ with a doping-independent slope [see Fig.~\ref{Fig4}]. Least-square linear fits of $ \Delta$ vs $H_{\textrm{ext}}$ in this higher field region give $g\mu_B/k_B \approx 1.4$, which corresponds to $g= 2$ as expected for the ground state $^8S_{7/2}$ of Eu. 

\section{Conclusions}
We analyzed the low-temperature specific heat data in order to investigate the interaction between magnetism and superconductivity and reveal the effect of magnetic field on the nature of the superconducting and antiferromagnetic orders in the Pr$_{1-x}$Eu$_x$Pt$_4$Ge$_{12}$ filled skutterudite system. Our data show the presence of short range AFM correlations between Eu ions under the superconducting dome for $ x \leq 0.50$. These short range AFM correlations produce a local internal magnetic field, which lifts the eight-fold degeneracy of the Eu ground state and gives rise to a Schottky peak in heat capacity that shifts to higher temperature with increasing $H_{\textrm{ext}}$. The low temperature heat capacity data can be fitted with the sum of a superconducting/normal state term and a Schottky term. In low values of $H_{\textrm{ext}}$, the internal and external magnetic fields are comparable, hence  the Schottky gap shows a non-linear dependence on $H_{\textrm{ext}}$. In high magnetic field, the applied magnetic field aligns the internal moment in its direction, hence the Schottky anomaly increases linearly with $H_{\textrm{ext}}$ with a doping-independent slope, as normally expected.  

\section{Acknowledgments} 
This work was supported by the National Science Foundation grants DMR-1505826 and DMR-1506547 at KSU and by the
US Department of Energy, Office of Basic Energy Sciences, Division of Materials Sciences and Engineering, under Grant
No. DE-FG02-04ER46105 at UCSD. The work of M. D. and P. S. was financially supported in part by the U.S. Department of Energy, Office of Basic Energy 
Sciences under Award No. DE-SC0016481. 
\bibliography{PrEuPtGe}

\end{document}